\documentstyle[psfig]{kapproc} 

\kluwerbib

\startauthorindex

\begin{document}

\articletitle{SEDs of Flared Dust Disks}

\articlesubtitle{Radiation Transfer Model Versus 2-Layer Model}

\author{Michaela Kraus}
\affil{Sterrenkundig Instituut, Universiteit Utrecht, \\
Princetonplein 5, NL--3584 CC Utrecht, The Netherlands}
\email{M.Kraus@phys.uu.nl}



\anxx{Michaela Kraus}

\begin{abstract}
A radiation transfer model for the top-layer of a flared dust disk around a
low mass star is presented. The disk structure,
temperature distribution and spectral energy distribution are compared with
the results of the analytical two-layer model of Chiang \& Goldreich. We find 
that both, a proper iterative calculation of the main parameters and proper 
radiation transfer calculations, are important to not overestimate the disk 
structure and its emission especially at long wavelengths.
\end{abstract}

\section*{Introduction}

Many T Tauri stars are observed with a flat spectral energy distribution 
(SED) in the near infrared (e.g. Beckwith et al. 1990). This is in 
contradiction with the standard disk which is usually modeled as a flat
blackbody. The flat spectrum in the near infrared hints to an additional hot 
dust component in the circumstellar region of young stars. Several models 
have been proposed, e.g. a dusty halo (e.g. Miroshnichenko et al. 1999) or 
a dust shell beyond the outer edge of the disk (e.g. Gregorio-Hetem \& Hetem 
2002). Both models can fit observed SEDs. However, it 
is not clear how such a configuration of the circumstellar dust should form.

We are dealing here with a different location of the hot dust component, namely
the surface of the circumstellar disk. For the surface to become much hotter 
than the interior, it must be inclined with respect to the infalling stellar
radiation, i.e. the disk must be flared. Such a flaring can happen naturally
when the disk is in hydrostatic equilibrium in $z$-direction with a temperature 
distribution that falls off more slowly than $r^{-1}$ (Kenyon \& Hartmann 1987).

Chiang \& Goldreich (1997, hereafter CG) developped an analytical two-layer 
model. Their flared dust disk contains an optically thick isothermal 
mid-layer and a superheated surface layer. CG provide simple analytical 
expressions to calculate the SEDs which are indeed flat in the near infrared,
and their model is used by many people to fit the SEDs of young stellar
objects.

Our intention was to study the reliability of these analytical two-layer models.
We developped a radiation transfer code for the upper disk part to 
calculate the parameters that determine the disk structure and temperature 
distribution self-consistently, and we compare our results with the CG model. 
Note that we compare our results to the original analytical CG model. A 
comparison between an improved two-layer model with a radiation transfer
model has recently been done by Dullemond \& Natta (2003).

\section{Description of the dust disk model}\label{sec3}

Our disk consists of an optically thick ($\tau_{\rm V}^{\rm mid} \gg 1$)
mid-layer which is isothermal in $z$-direction. The mid-layer is sandwiched by 
two top-layers which are marginally optically thick at visual wavelengths but 
still optically thin at IR wavelengths, i.e. $\tau_{\rm V}^{\rm
top} \ge 1 \quad$ and $\quad \tau_{\rm IR}^{\rm top} \le 1$. Consequently,
the infalling stellar light is completely absorbed within the top-layer. The 
dust particles re-radiate the energy at IR wavelengths, so half
the redistributed energy leaves the disk into space, and the other half
penetrates the mid-layer and heats it.
As a third distinct region we define the disk photosphere which encloses the
uppermost part of the top-layer and whose location is defined by the parameter 
$h$ (see Sect.\,\ref{atmosphere}). 

Further, we make the following assumptions: the disk is in hydrostatic 
equilibrium in $z$-direction, gas and dust are well mixed throughout the disk, 
and in the mid-layer gas (g) and dust (d) are in thermal equilibrium at the 
same temperature $T_{\rm g} = T_{\rm d} = T_{\rm mid}$. 
The emission and absorption of the gas component of the disk is ignored.

We restrict the discussion to passive disks, which means that the only heating 
source is the star which illuminates the disk surface. In addition, our (gas) 
disk extends down to the stellar surface with no inner hole. The existence of 
such a hole would lead to a puffed-up inner rim and self-shadowing effects of 
the disk (see Dullemond et al. 2001 and Dullemond 2002).

\subsection{Stellar illumination of the disk surface}

For a razor-thin passive disk around a star with effective temperature $T_{*}$ and radius
$R_{*}$ the monochromatic flux entering the disk perpendicular to the surface at 
distance $r$ from the star is
\begin{equation}\label{f_real}
f^{\perp}_{\nu} = B_{\nu}(T_{*})\left[\arcsin\frac{R_{*}}{r} -
\frac{R_{*}}{r}\sqrt{1-\left(\frac{R_{*}}{r}\right)^{2}}~\right]
\end{equation}  
where the star is assumed to emit a Planck spectrum. This flux can be 
parametrized in terms of the solid angle $\Omega$ under which the star is seen 
from a disk surface element at distance $r$ and the so-called grazing 
angle $\alpha_{\rm gr}$, i.e. the mean angle of incidence of the stellar flux 
\begin{equation}\label{alfa_def}
f^{\perp}_{\nu} = \alpha_{\rm gr} \Omega B_{\nu}(T_{*})
\end{equation}
The grazing angle of a razor-thin disk is therefore
\begin{equation}\label{alfa_razor}
\alpha_{\rm gr}^{\rm razor}  =  \frac{1}{\Omega}\left[\arcsin\left(
\frac{R_{*}}{r}\right) - \frac{R_{*}}{r}\,\sqrt{1-\left(\frac{R_{*}}
{r}\right)^{2}}~\right]
\end{equation}
which for large distances reduces to the handy formula used by CG
\begin{equation}\label{alfa_razor_red}
\alpha_{\rm gr}^{\rm razor} \stackrel{r\gg R_{*}}{\longrightarrow}~\frac{4}{3\pi}
\frac{R_{*}}{r} \simeq 0.4 \frac{R_{*}}{r}
\end{equation}
This relation also holds for a wedge-shaped disk as long as its opening angle is small. 

\begin{figure}[t]
\narrowcaption{Sketch of a flared disk. The flaring term of the grazing angle 
is indicated. The inner radius of the dust disk is $r_{0}$.}
\vspace{-1.7cm}
\psfig{file=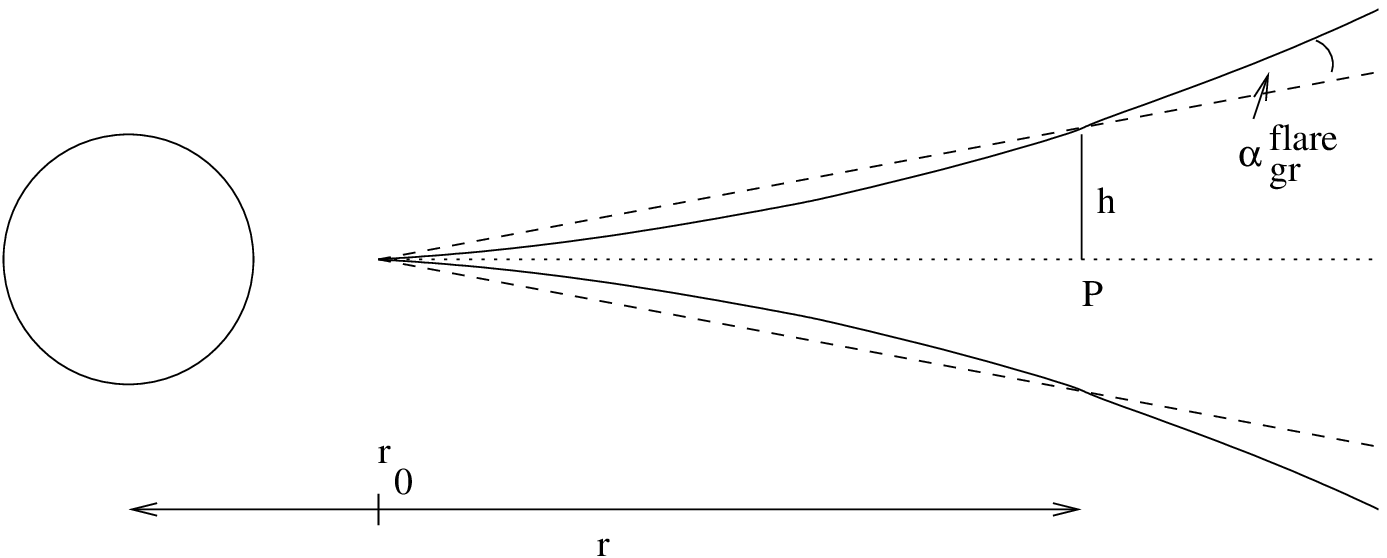,width=18pc}
\label{flare}
\end{figure}

In a flared disk the surface is curved and the grazing angle increases with distance
from the star (see the the solid lines in Fig.\,\ref{flare}). Therefore, the flared
disk intercepts more stellar light and is heated more efficiently.
The grazing angle of the flared disk is by the amount $\alpha_{\rm gr}^{\rm flare}$ 
greater than for a flat disk (Fig.\,\ref{flare}), and the total grazing angle becomes 
simply $\alpha_{\rm gr} = \alpha_{\rm gr}^{\rm razor} + \alpha_{\rm gr}^{\rm flare}$
This flaring term of the grazing angle, $\alpha_{\rm gr}^{\rm flare}$, can be computed from
\begin{equation}\label{alfa_h}
\alpha_{\rm gr}^{\rm flare}  =  \arctan \frac{{\rm d}h}{{\rm d}r} -
\arctan\frac{h}{r}\simeq r \frac{\rm d}{{\rm d}r}
\left(\frac{h}{r}\right)
\end{equation}
With this new grazing angle, which is a function of the location $h$ of the photosphere,
we can calculate the flux penetrating the flared disk in $z$-direction.

\subsection{The location of the disk photosphere}\label{atmosphere}

The photosphere encloses the uppermost part of the top-layer. Its onset is described
by the parameter $h$, which is defined as the height $z$ above the mid-plane where
the visual optical depth along the direction of the infalling stellar light, i.e.
along the grazing angle, equals 1. This leads to a vertical optical depth of
\begin{equation}\label{tau_perp}
\tau^{\perp}_{\rm V} = \sin \alpha_{\rm gr}\,.
\end{equation}
Since the disk is assumed to be in hydrostatic equilibrium, $\tau^{\perp}_{\rm V}$
can be calculated
\begin{equation}\label{tau2}
\tau_{\rm V}^{\perp} = \kappa_{\rm V}\rho_{0}(r)\int\limits_{h}^{\infty}e^{-
\frac{z^{2}}{2H^{2}}}\,dz
\end{equation}
where $\rho_{0}$ is the density in the mid-plane and $H$ is the scale height of the disk
given by
\begin{equation}\label{scale}
H = \sqrt{\frac{kT_{\rm d}}{GM_{*}\mu}}~r^{3/2}
\end{equation}
Here, $M_{*}$ and $\mu$ are the stellar mass and the mean molecular weight.

From equaling Eqs.\,(\ref{tau_perp}) and (\ref{tau2}) $h$ can be determined, but it
is a function of temperature and grazing angle.

\subsection{The temperature of the mid-layer}

The downwards and upwards directed fluxes, $F^{\downarrow}$ and $F^{\uparrow}$, 
are in equilibrium throughout the passive disk. 
We can calculate these two fluxes explicitely
at the boundary between the isothermal mid-layer and the top-layer. 
We assume that half the incident stellar flux is re-radiated into space and
the other half penetrates the mid-layer, leading to
\begin{equation}\label{fdown}
F^{\downarrow} = \frac{1}{2}\,f^{\perp} = \frac{1}{2}\int \alpha_{\rm gr}
\Omega B_{\nu}(T_{*})\,d\nu
\end{equation}
The flux leaving the mid-layer in upward direction is given by
\begin{equation}\label{fup}
F^{\uparrow} = 2\pi\int\!\!\!\int B_{\nu}(T_{\rm mid})\left(
1-e^{-\tau_{\nu}/\mu}\right)\mu\,d\mu\,d\nu
\end{equation}
where $T_{\rm mid}$ is the isothermal temperature of the mid-layer, and we set $\mu =
\cos\theta$ with $\theta$ as the angle measured from the $z$-axis. The visual optical
depth of the mid-layer in vertical direction can be written in the form
\begin{equation}
\tau_{\rm V}^{\rm mid}(r) = \tau_{\rm V}^{\rm mid}(r_{0})\left(\frac{r}{r_{0}}\right)^{-s}
\end{equation}
with the visual optical depth at the inner edge, $\tau_{\rm V}^{\rm 
mid}(r_{0})$, and the exponent $s$ as free parameters. The temperature of the 
mid-layer, $T_{\rm mid}$, follows from equaling Eqs.\,(\ref{fup}) and 
(\ref{fdown}). This temperature is, however, a function of the grazing angle.

We have now three important parameters, $\alpha_{\rm gr}, h,$ and $T_{\rm mid}$,
but none of them can be computed independently, instead they must be calculated
iteratively.

\subsection{Radiation transfer within the top-layer}\label{radtransfer}

We have to specify the visual optical depth in the top-layer in $z$-direction,
$\tau^{\rm top}_{\rm V}$.  It should be high enough so that first, the infalling
stellar radiation is completely absorbed and second, the dust grains at the 
bottom of the top--layer reach a temperature close to that of the mid-layer in
order to guarantee a smooth transition.  On the other hand, $\tau^{\rm top}_{\rm
V}$ must be small enough to allow the dust emission, which occurs at infrared
wavelengths, to escape the top-layer. 

The radiation transfer equation of a plane-parallel slab,
\begin{equation}\label{strahl}
I_{\nu}(\mu,\tau_{\nu}) = I_{0}\,e^{\tau_{\nu}/\mu} - \int
S_{\nu}(t)e^{-(t-\tau_{\nu})/\mu}\,\frac{dt}{\mu}
\end{equation}
with the source function $S_{\nu}$, the incident intensity $I_{0}$ and
$\mu = \cos\theta$ is split
into up-streams, $I^{+}_{\nu}$, that penetrate the top-layer from the 
mid-layer, and down-streams, $I^{-}_{\nu}$, that cross the top-layer 
starting from the surface. The incident intensity is either the 
stellar radiation for downwards directed streams (but only for angles 
$\theta$ under which the star can be seen, else it is zero), or the 
emission of the mid-layer, $B_\nu(T_{\rm mid})(1-e^{-\tau_\nu^{\rm mid}/\mu})$,
for the upwards directed streams.  

With the help of the Feautrier parameters
\begin{equation}
u_{\nu} = \frac{1}{2}\,(I^{+}_{\nu} + I^{-}_{\nu}) \qquad \textrm{and} 
\qquad  v_{\nu} = \frac{1}{2}\,(I^{+}_{\nu} - I^{-}_{\nu})
\end{equation}
the mean intensity $J_{\nu}$ becomes
\begin{equation}
J_{\nu} = J_{\nu}(\tau_{\nu}) = \int u_{\nu}(\mu,\tau_{\nu})\,d\mu
\end{equation}
which is needed to calculate the source function and the emission of the
grains. The source function itself determines the up- and down-streams of
the intensity; the radiation field calculation must therefore be iterated. 

\begin{figure}[th]
\sidebyside
{\centerline{\psfig{file=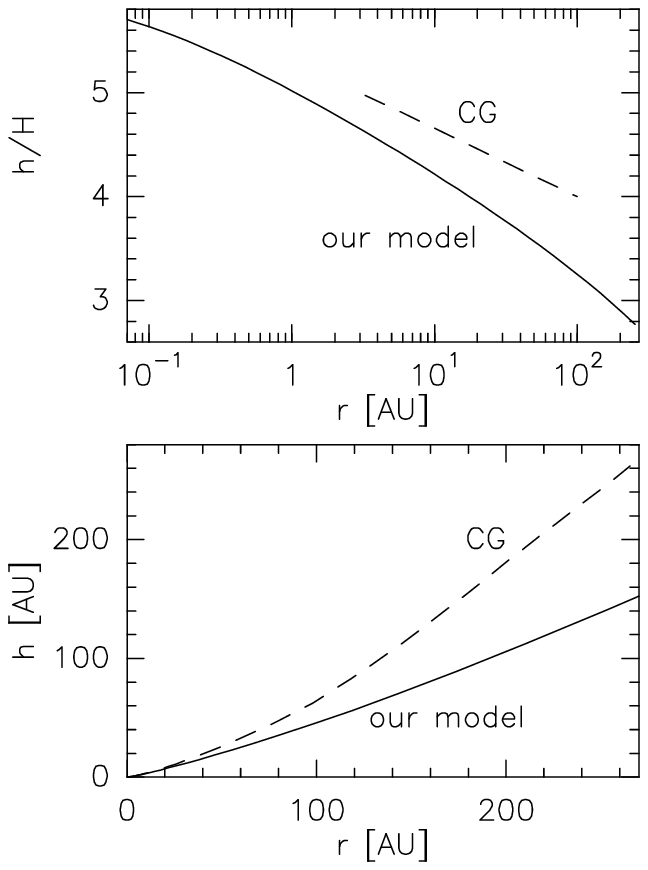,width=13pc}}
\caption{Drop of the parameter $h/H$ (top, see text), and
the height $h$ of the photosphere (bottom).}}
{\centerline{\psfig{file=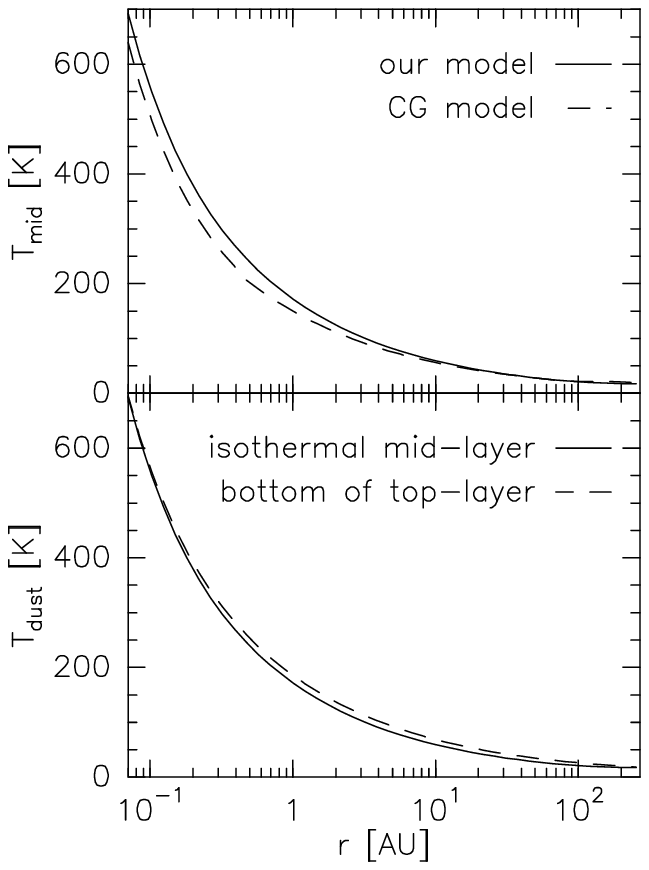,width=13pc}}
\caption{Temperature of the mid-layer (top), and comparison of
the top- and mid-layer temperature at the boundary (bottom).}}
\end{figure}

Finally, the temperature of the grains at each location in the 
top-layer follows from balancing their absorption from the surrounding 
radiation field, $J_{\nu}$, and their emission at their equilibrium 
temperature, $T_{\rm d}$.

\section{Results}

Since we want to compare our results with those of CG we took the same 
parameters as they did: for the star, $T_{*} = 4000$\,K, $L_{*} = 
1.44$\,L$_{\odot}$, $R_{*} = 2.5$\,R$_{\odot}$, $M_{*} = 0.5$\,M$_{\odot}$; 
for the disk, $r_{\rm in} \simeq 0.07$\,AU, $r_{\rm out} = 270$\,AU, 
$\tau_{\rm V}^{\rm mid}(r) \simeq 4 \cdot 10^{5}r_{\rm AU}^{-1.5}$; for the 
dust grains, spheres of radius $a = 0.1\,\mu$m with mass density $\rho_{\rm 
d} = 2$\,g\,cm$^{-3}$ and absorption efficiency $Q = 1$ for $\lambda \le 2\pi 
a$ and $Q \simeq 2\pi a / \lambda$ for $\lambda \ge 2\pi a$.  The latter 
implies an absorption coefficient $\kappa_{\rm V} \simeq 4 \times 
10^4$\,cm$^{2}$ per gram of dust.  Scattering is absent and we use $\tau_{\rm 
V}^{\rm top} = 3$. The disk is seen face-on.

\begin{figure}[th]
\sidebyside
{\centerline{\psfig{file=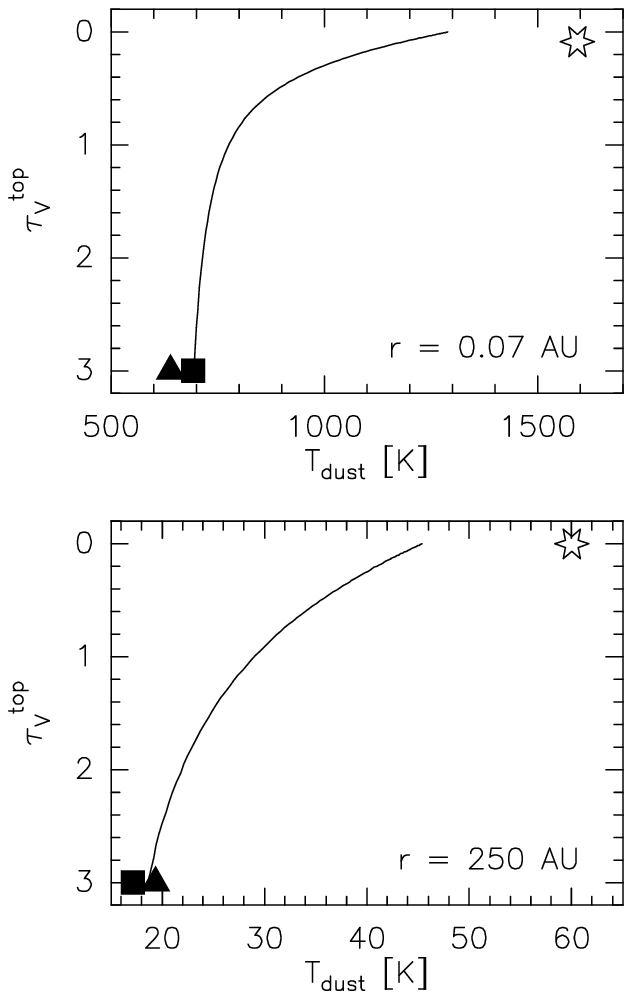,width=13pc}}
\caption{Temperature distribution within the top-layer at $r_{o}$ (top)
and at $r = 250$\,AU (bottom). The star is the CG value, square and triangle
are our and the CG mid-layer value.}}
{\centerline{\psfig{file=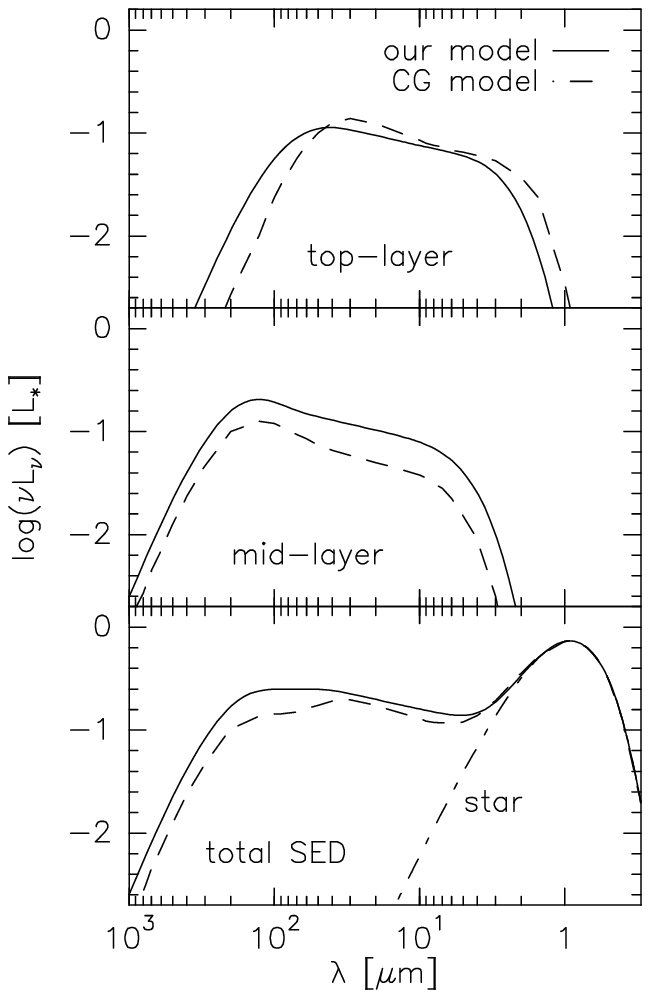,width=13pc}}
\caption{Comparison between CG and our results for the top-layer (top),
mid-layer (middle) and total SED (bottom).}}
\end{figure}

We first compare our height $h$ of the photosphere (Fig.\,2) with the CG 
results. The ratio $h/r$ which determines $\alpha_{\rm 
gr}$ can be written in terms of the scale height $H$
\begin{equation}
\frac{h}{r} = \frac{h}{H}\,\frac{H}{r}
\end{equation}
According to CG the ratio $h/H$ drops from 5 at 3\,AU to 4 at 100\,AU but they
use $h/H = 4$ throughout their calculations which leads to an extremely thick
flared disk with $h(r=270$\,AU$) = 270$\,AU. Our calculations show, however, 
that $h/H$ drops from about 6 at $r_{0}$ to $\leq 2.8$ at the outer edge,
and our disk stays much flatter at large distances. In a recent work Chiang
et al. (2001) corrected their old results for the disk thickness and their 
corrected photosphere almost agrees with our results.

The temperature in the mid-layer, $T_{\rm mid}$ (Fig.\,3), is not so different
in the two models. Our mid-layer is slightly hotter especially in the inner
parts ($r\leq 20$\,AU) but comparable with CG for larger distances.  
Our temperature calculations in the top-layer result in a temperature
gradient between the very hot surface and the bottom of the top-layer.
In Fig.\,4 we compare our temperature distribution with the CG surface 
temperature as well as with our and the CG mid-layer temperatures at two 
different distances. We cannot reproduce the very hot surface temperature of 
CG, but on the bottom of the top-layer (at $\tau_{\rm V}^{\rm top} = 3$) the 
temperature approaches everywhere the value of the mid-layer (see bottom
panel of Fig.\,3). 

The differences in the disk and temperature structure between the CG model and 
our more detailed radiation transfer calculations lead of course also to
differences in the SED (Fig.\,5). The emission from the CG surface layer
which is much hotter than our top-layer is shifted to shorter wavelengths.
The emission of the CG mid-layer comes from a narrower wavelength 
region than in our model and is too low. The total SED of CG therefore 
deviates from our results especially in the long wavelength region ($\lambda 
\geq 30\mu$m).

\section{Conclusions}

A radiation ransfer model for the upper disk part of a flared disk is 
presented and the results are compared with the analytical two-layer model 
of CG. We show that the major parameters that determine the structure of 
the disk, $\alpha_{\rm gr}, h, T_{\rm mid}$, must be calculated iteratively.  
In addition we perform detailed radiation transfer calculations in the 
top-layer to find the temperature structure in the upper disk part. At the 
bottom of the top-layer the temperature agrees very good with the values of 
the mid-layer. The very hot surface temperature found by CG could not be 
reproduced, and our total SED deviates from the one of CG especially
in the long wavelength regime. Although the two-layer model gives handy 
formulae to calculate the SED of a flared dust disk, the results should only 
be used as a `first guess' for the structure and temperature distribution of
the disk and radiation transfer calculations should be performed for a 
better characterization of the circumstellar dust disk.  

\begin{acknowledgments}
I would like to thank Dr.~Endrik Kr\"{u}gel for many helpful discussions.
This work was supported by the German \emph{Deut\-sche 
For\-schungs\-ge\-mein\-schaft, DFG\/} grant number Kr~2163/2--1.
\end{acknowledgments}


\begin{chapthebibliography}{1}
\bibitem{beckwith}
        Beckwith, S.V.W., Sargent, A.I., Chini, R.S. \& G\"{u}sten, R.
        1990, AJ 99, 924
\bibitem{chiang}
        Chiang, E.I., \& Goldreich, P. 1997,
        ApJ 490, 368
\bibitem{chiang2}
        Chiang, E.I., Joung, M.K., Creech-Eakman, M.J., Qi, C.,
        Kessler, J.E., Blake, G.A. \& van Dishoeck, E.F. 2001,
        ApJ 547, 1077
\bibitem{dullemond02}
        Dullemond, C.P. 2002, A\&A 395, 853
\bibitem{dullemond03}
        Dullemond, C.P., \& Natta, A. 2003, A\&A {\it in press}
\bibitem{dullemond01}
        Dullemond, C.P., Dominik, C., \& Natta, A. 2001,
        ApJ 560, 957
\bibitem{hetem}
        Gregorio-Hetem, J. \& Hetem, A., Jr. 2002, 
        MNRAS 336, 197
\bibitem{kenyon}
        Kenyon, S.J., \& Hartmann, L. 1987,
        ApJ 322, 293
\bibitem{miro}
        Miroshnichenko, A., Ivezic, Z., Vinkovic, D. \& Elitzur, M. 1999,
        ApJ 520, L\,115
\end{chapthebibliography}
\end{document}